\documentclass[aps,prb,preprint,superscriptaddress]{revtex4}
\usepackage{times,xspace}
\usepackage{amsbsy,amssymb,amsmath,bm}
\usepackage{graphicx,color,epsfig,rotate}
\usepackage{fancyheadings}

\begin{document}
\title{DOMAIN SIZE DEPENDENCE OF PIEZOELECTRIC PROPERTIES OF FERROELECTRICS}
\author{Rajeev Ahluwalia}
\author{Turab Lookman}
\author{Avadh Saxena}
\affiliation{Theoretical Division, Los
Alamos
National Laboratory,
Los Alamos, New Mexico 87545 }
\author{Wenwu Cao}
\affiliation{Materials Research Institute and Department of Mathematics, 
Pennsylvania State University, University Park, Pennsylvania 16802 }
\date{\today}
\begin{abstract}
The domain size dependence of piezoelectric properties of
ferroelectrics is investigated using a continuum Ginzburg-Landau
model that incorporates the long-range elastic and electrostatic
interactions. Microstructures with desired domain sizes are
created by quenching from the paraelectric phase by biasing the initial
conditions. Three different two-dimensional microstructures with different
sizes of the $90^{o}$ domains are simulated. An electric field is
applied along the polar as well as non-polar directions and the
piezoelectric response is simulated as a function of domain size
for both cases. The simulations show that the piezoelectric
coefficients are enhanced by reducing the domain size, consistent with
recent experimental results of Wada and Tsurumi (Brit. Ceram. Trans. {\bf 103}, 93, 2004) on domain engineered $BaTiO_{3}$
single crystals. 
\end{abstract}
\maketitle
\section{INTRODUCTION}
Ferroelectrics are excellent  piezoelectric materials that can convert electrical
energy into mechanical energy and vice versa \cite{glass}. This
electromechanical property
arises due to the coupling of  spontaneous polarization with
lattice strain.
Many devices such as ultrasonic transducers and piezoelectric actuators make use of this
property\cite{uchino}. Recently, there has been considerable
interest in this field due to the observation of a giant
piezoelectric response if the applied field is along a non-polar
direction \cite{shrout,wada}.
 It is believed that this
``superpiezoelectric" response is due to the symmetry change caused
by a rotation of the polarization
 towards the direction of the applied field\cite{cohen}.
Domain configurations produced by the field applied in the non-polar direction
are termed {\it engineered domains}. There are also a large number
of domain walls between the degenerate variants which affect the
piezoelectric property. In a recent paper, Wada and Tsurumi \cite{dmsize}
studied the dependence of the piezoelectric properties of domain
engineered $BaTi{O_3}$ single crystals as a function of domain
size. Engineered domain configurations with a range of domain
sizes were synthesized. The study revealed that piezoelectricity
is enhanced for domain engineered crystals with small domain sizes
(or high domain wall density). Thus, domain walls influence the
piezoelectric properties and it is important to compute the
contribution of the domain walls to the piezoelectric response.

Electromechanical properties of ferroelectrics have been studied
theoretically using first-principle calculations
\cite{david,cohen,nasai}. A continuum Landau theory describing a
single domain or homogeneous state has been used to study the
electromechanical properties of $BaTi{O_3}$ as a function of
temperature and electric field direction \cite{bell}. Although
such calculations provide valuable insights into the physics of
the polarization-strain coupling, they do not describe
inhomogeneities due to domains and domain walls. Recently, we
studied the piezoelectric properties of domain engineered two-dimensional 
($2D$) ferroelectrics using the time-dependent Ginzburg Landau (TDGL)
theory \cite{rapl}. The  important conclusion from our
simulations was the role played by the domain walls in nucleating
an electric field induced structural transition if an electric
field is applied along a non-polar direction. We showed that the
field induced transition occurred at lower electric fields for a
multi-domain state, compared to an analogous situation for a
single domain state. To understand the recent experimental results of
Wada and Tsurumi \cite{dmsize} that show piezoelectric enhancement at small
domain sizes, we extend in this paper the TDGL model to
investigate the dependence of piezoelectricity on  the size of the
$90^{o}$ domains in the system and the domain wall density.
Unlike Ref. [10] where the domain
microstructure was obtained by quenching from the paraelectric
phase with random initial conditions, here we  create domain
structures with desired sizes by appropriately biasing the initial
conditions. This procedure allows us to obtain domain
microstructures with a range of domain sizes. The size dependence
is studied for the case with the electric field along a polar axis as
well as that with the field along a non-polar direction.

The paper is organized as follows. In Sec. II, we describe the
model in detail. Section III describes our simulations for the
case of the  electric field applied along a polar axis. In Sec. IV,
we discuss the case in which the  electric field is applied along a non-polar
direction. We conclude in Sec. V with a summary and
discussion.
\section{THE MODEL}
The calculations are based on a time-dependent Ginzburg-Landau
model \cite{sagala,lqc,rawc} with long-range elastic and electrostatic
effects. We restrict ourselves to a $2D$ ferroelectric transition
to illustrate the basic principles  and use parameters from a
model for $BaTi{O_3}$ in our calculations \cite{bell}. The
free-energy functional for a $2D$ ferroelectric system is written
as
$F=F_{l} + F_{em} +F_{es}$.
Here $F_{l}$ is the local free energy \cite{bell} that describes the
ferroelectric transformation and is given by 
\begin{eqnarray}
&F_{l}&=\int d\vec{r}\bigg\{{\alpha_1}(P_x^2+P_y^2)
+{\alpha_{11}}(P_x^4+P_y^4)
+{\alpha_{12}}P_x^2P_y^2
+{\alpha_{111}}(P_x^6+P_y^6)
+{\alpha_{112}}(P_x^2P_y^4\nonumber \\
&+&P_x^4P_y^2)
-{E_x}{P_x}-{E_y}{P_y}
+{{g_1}\over{2}}( P_{x,x}^2
+P_{y,y}^2)
+
{{g_2}\over{2}}(P_{x,y}^2+
P_{y,x}^2)
+
{g_3}P_{x,x}P_{y,y}
\bigg\} ,
\end{eqnarray}
 where $P_x$ and $P_y$ are the polarization components.
The free energy coefficients $\alpha_1, \alpha_{11},...,\alpha_{112}$
determine the ferroelectric phase and the gradient
coefficients $g_1$, $g_2$ and $g_3$ are a measure of domain wall energies. $E_x$ and $E_y$ are the components of an external electric field.
Elastic properties are studied by
using the strains $\eta_1=\eta_{xx}+\eta_{yy}$
, $\eta_2=\eta_{xx}-\eta_{yy}$ and
$\eta_3=\eta_{xy}$, where $\eta_{ij}$ is the
linearized strain tensor defined as
 $\eta_{ij}=(u_{i,j}+u_{j,i})/2$ $(i,j=x,y)$, $u_i$ being the components of the displacement
vector.
The electromechanical coupling is described in terms of these strain variables
with the free energy
\begin{equation}
F_{em}=\lambda\int d\vec{r}~\left[\{\eta_1-{Q_1}({P_x}^2+{P_y}^2)\}^2+
\{\eta_2-{Q_2}({P_x}^2-{P_y}^2)\}^2
+\{\eta_3-{Q_3}{P_x}{P_y}\}^2\right].
\end{equation}
Here $Q_1$, $Q_2$ and $Q_3$ are obtained from the electrostrictive
constants of the material with $Q_1=Q_{11}+Q_{12}$,
$Q_2=Q_{11}-Q_{12}$ and $Q_3=Q_{44}$ (electrostrictive constants
describe coupling between strains and polarization, that is,
$\eta_{xx}=Q_{11}{P_x}^2+Q_{12}{P_y}^2$,
$\eta_{yy}=Q_{11}{P_y}^2+Q_{12}{P_x}^2$ and
$\eta_{xy}=Q_{44}{P_x}{P_y}$). Notice that the free energy $F_{em}$
vanishes for a homogeneous state since the homogeneous strains in
equilibrium are given by ${\eta_1}^{e}={Q_1}({P_x}^2+{P_y}^2)$,
${\eta_2}^{e}={Q_2}({P_x}^2-{P_y}^2)$, and
${\eta_3}^{e}={Q_3}{P_x}{P_y}$. However, this free energy does not
vanish for an inhomogeneous state. For an inhomogeneous state, the
strains $\eta_1$, $\eta_2$ and $\eta_3$ are related to each other
by the elastic compatibility constraint \cite{love}
\begin{equation}
{{\nabla}^2}{\eta_1}-({{\partial^2}\over{\partial x^2}}
-{{\partial^2}\over{\partial y^2}}){\eta_2}-
{{\partial^2}\over{ {\partial x}{\partial y} }}{\eta_3}=0.
\end{equation}
Using this relation, the strain $\eta_1$ can be eliminated from $F_{em}$
resulting in a nonlocal interaction between the strains involving $\eta_2$ and $\eta_3$. Using the equilibrium strains defined by
${\eta_2}^{e}$ and
${\eta_3}^{e}$, the electromechanical free energy can be written as
\begin{equation}
F_{em}=\lambda\int d \vec{k}~\left|{C_2(\vec{k})}\Gamma_2(\vec{k})
+{C_3(\vec{k})}\Gamma_3(\vec{k})
-\Gamma_1(\vec{k})\right|^2,
\end{equation}
 where the homogeneous state corresponding to the $\vec{k}=0$ mode has been excluded from the above integral.
The constant $\lambda$ is the strength of this nonlocal interaction and hence it influences the underlying
microstructure.
The quantities $\Gamma_1(\vec{k})$, $\Gamma_2(\vec{k})$ and $\Gamma_3(\vec{k})$ are respectively the Fourier
transforms of $Q_1({P_x}^2+{P_y}^2)$
, $Q_2({P_x}^2-{P_y}^2)$ and
$Q_3{P_x}{P_y}$;
$C_{2}=({k_x}^2-{k_y}^2)/
({k_x}^2+{k_y}^2)$ and
 $C_{3}={k_x}{k_y}/
({k_x}^2+{k_y}^2)$ are the orientation dependent kernels. The
electrostatic contribution to the free energy is calculated by
considering the depolarization energy \cite{brat}
\begin{equation}
F_{es}=-\mu\int d\vec{r}
~\left\{\vec{E_d}\cdot\vec{P}
+{\epsilon_0}({\vec{E_d}\cdot\vec{E_d}}/{2})\right\},
\end{equation}
where
$\vec{E_d}$ is the internal depolarization field due to the
dipoles and $\mu$ is the strength of this interaction. The field
$\vec{E_d}$ can be calculated from an underlying potential using
$\vec{E_d}=-\vec{\nabla}\phi$. If we assume that there is no free
charge in the system, then $\vec{\nabla}\cdot\vec{D}=0$, where
$\vec{D}$ is the electric displacement vector
 defined by $\vec{D}={\epsilon_0}\vec{E_d}+\vec{P}$. This equation gives rise to the
constraint
$-{\epsilon_0}{\nabla}^2{\phi}+\vec{\nabla}\cdot\vec{P}=0$. The
potential $\phi$ is eliminated from the free energy $F_{es}$ using
the above constraint to express $F_{es}$ in Fourier space as
\begin{equation}
F_{es}=\frac{\mu}{2\epsilon_0}\int d\vec{k}~\left|\hat{k_x}{P_x}(\vec{k})
+\hat{k_y}{P_y}(\vec{k})
\right|^2.
\end{equation}
The above integral excludes the homogeneous $\vec{k}=0$ mode
which means that the homogeneous depolarization field due to
surface charges has been neglected. The total energy is defined by 
$F=F_{l}+F_{em}+F_{es}$ with two additional constants, i.e.
$\lambda$ and $\mu$ are essential for the description of
multi-domain states.

The dynamics of the polarization fields is given by the relaxational
time-dependent Ginzburg-Landau equations
\begin{equation}
{{\partial P_i}\over{\partial t}}=-
{\gamma}{{\delta F}\over{\delta P_i}},
\end{equation}
where $\gamma$ is a dissipation coefficient and $i=x,y$ represents the polarization components.
We first introduce rescaled variables defined with 
$u=P_x/{P_0}$,
$v=P_y/{P_0}$, $\vec{\zeta}=\vec{r}/\delta$ and $t^*=\gamma|\alpha_1(T_0)|{t}$, where
$T_0$ is a fixed temperature.
In this work, we use the parameters \cite{bell} for $BaTi{O_3}$ for the local part of the free energy
$F_{l}$. The parameters which can be dependent on the temperature $T$ are:
$\alpha_1=3.34\times{10^5}(T-381)$ VmC$^{-1}$,
$\alpha_{11}=4.69\times{10^6}(T-393) -2.02\times{10^8}$
Vm$^{5}$C$^{-3}$,
$\alpha_{111}=-5.52\times{10^7}(T-393)
+2.76\times{10^9}$ Vm$^{9}$C$^{-5}$,
$\alpha_{12}=3.23\times{10^8}$ Vm$^{5}$C$^{-3}$ and
$\alpha_{112}=4.47\times{10^9}$ Vm$^{9}$C$^{-5}$. The electrostrictive
constants are given as $Q_{11}=0.11$ m$^{4}$C$^{-2}$,
$Q_{12}=-0.045$ m$^{4}$C$^{-2}$ and
$Q_{44}=0.029$ m$^{4}$C$^{-2}$.
We assume that the coefficients $g_1=g_2=g_3=g$ and use the value
$g=0.025\times{10^{-7}}$ Vm$^3$/C quoted in the literature \cite{size}.
To calculate the rescaled quantities, we use $T_0=298K$, $P_0=0.26$
 Cm$^{-2}$ and
$\delta\sim 6.7$ nm. The values chosen for the long-range parameters are
$\lambda=0.25|\alpha_1(T_0)|/{P_0}^2$ and
$\mu=20{\epsilon_0}|\alpha_1(T_0)|$.  All results presented below are 
expressed in terms of the rescaled time $t^\ast$.  
\section{SIMULATIONS}
The time-dependent Ginzburg-Landau model with the above rescaled
parameters is used to simulate the domain patterns and
electromechanical properties. The equations are discretized on a
$128\times 128$ grid with the Euler scheme using periodic boundary
conditions. For the length rescaling factor $\delta \sim 6.7$ nm,
this discretization corresponds to a system of size $\sim 0.85$
$\mu$m $\times 0.85$ $\mu$m. We simulate the properties of this
$2D$ model at $T=298 K$. At this temperature, the minima of the
free energy $F_l$ define a rectangular ferroelectric phase with
the four degenerate states $(\pm 0.26, 0)$ Cm$^{-2}$ and $(0, \pm
0.26)$ Cm$^{-2}$. Since we want to study the domain size
dependence of properties, we create domain structures with
required domain size instead of letting the domain structure form
after a quench from the paraelectric phase. This is achieved by choosing
initial conditions based on the following procedure. We consider a function
\begin{equation}
R(x,y)=\cos\left({{N\pi(x+y)}\over{128\delta}}\right).
\end{equation}
The initial conditions are set up by 
\begin{eqnarray}
{P_x}(x,y)=P_{0},~~ {P_y}(x,y)=0,~~~~~~~  R(x,y) >0 \nonumber\\
{P_x}(x,y)=0,~~ {P_y}(x,y)={P_0}.~~~~~~~  R(x,y) <0 
\end{eqnarray}  
These initial conditions ensure that multi-domain states with head to tail domain walls 
oriented along [11] are formed. 
The above initial conditions also ensure that only two of the four
variants with head to tail domain walls are formed  
in the multi-domain. 
 The quantity $N$ controls the number of domain walls and hence the domain size of the resulting
 microstructure.
We consider the cases $N=2,4,10$
corressponding respectively to $90^{o}$ domain patterns with mean
domain sizes $L_{0} \sim 0.3 {\mu}m, 0.15 \mu m, 0.06\mu m$. 
The top left snapshots in Figs. 1, 2 and 3 represent the prepared 
zero field multi-domain states for 
$L_{0} \sim 0.3 {\mu}m, 0.15 \mu m$ and $0.06\mu m$, respectively. 
These were obtained by solving Eqs. (7) for a time interval
$t^{*}=100 $ using the initial conditions given by Eqs. (8) and (9).
A close look at the 
local dipoles within the domains shows that 
the polarization vectors are $(P_0,\Delta)$ and
 $(\Delta, P_0)$, unlike the ideal single crystals which are described 
by $(P_0,0)$ or $(0,P_0)$.  This means that within the domains, the polarization vectors are slightly rotated compared to the single crystals.  
As the domain size becomes smaller, the quantity $\Delta$ increases and the polarization 
vectors within the domains get increasingly rotated from the 
ideal [10] and [01] directions. This rotation is  very strongly observed for the smallest domain size 
$L_0 \sim 0.06 \mu m$ (Fig. 3(a)).  
This is due to the fact that the domain walls are closely spaced
and the length of the diffuse interfaces is comparable to the domain
width.

To simulate the effect of an external electric field, the
evolution equations are solved  with a varying $\vec{E}$. We
consider two cases: (A) Field applied along the [01] direction, 
$\vec{E}=(0,E_{0})$, corresponding to the polar direction. (B)
Field applied along the [11] direction, 
$\vec{E}=({{E_{0}}\over{\sqrt{2}}},{{E_{0}}\over{\sqrt{2}}})$,
corresponding to a non-polar direction.
\subsection{FIELD APPLIED ALONG A POLAR DIRECTION}
We first study the traditional scenario when the electric field is
applied along one of the polar directions. In the present
simulations, we apply the field along the [01]
direction which is
a polar direction. The field is applied quasi-statically, i.e. in fixed
increments of $\Delta E_{[01]}= 0.92 kV/cm$ and we let the system relax for
$t^{*}=100$ time steps after each change.  Since [01] is a polar direction for the
parameters used in the present simulations, the state
$(0,0.26)$Cm$^{-2}$ is favored. Figures 1, 2 and 3 show the
electric field induced domain evolution for domain patterns with
mean domain sizes $L_{0} \sim 0.3 \mu m, 0.15 \mu m$ and 
$ 0.06 \mu m$, respectively. It can
be seen that domains aligned along the [10] direction switch towards
the [01] direction thereby forming a single domain state for all
the three cases. A comparison of the evolution in Figs. $1, 2$
and $3$ shows that a 
single domain state is established at smaller electric fields for
domain patterns with a larger number of
domain walls (or smaller domain sizes). For example, for the
smallest domain size $L_{0}\sim 0.06 \mu m$ (Fig. 3), the single domain
is established at an electric field $E_{[01]} \sim 2.5$ kV/cm
(this electric field is much smaller than the electric field
required to create single domain states for the domain patterns with
$L_{0} \sim 0.3 \mu m, 0.15 \mu m$). 
Figures 4(a) and 4(b) show the variation of the average
polarization with the applied field for the evolution depicted
in Figs. $1$, $2$ and $3$. For comparison, we also show the
polarization vs. electric field response of a single domain state
polarized along [01], i.e. $\vec{P}=(0,0.26)$ Cm$^{-2}$, when
an electric field is applied along the [01] direction. Figure 4(a)
shows the evolution of the [10] component of the average
polarization $P_{[10]}$. At zero field, for the three cases shown
in Figs. $1,2$ and $3$, $P_{[10]}$ has a non-zero value due to 
coexisting domains and domain walls. Interestingly, this average
value increases with decreasing the mean domain size.  This increase 
can be attributed to the rotation of the dipoles within the domains. 
As discussed earlier, the polarization vectors within the domains are 
given by 
$\vec{P}=({P_0},{\Delta})$ 
or $\vec{P}=({\Delta},{P_0})$. 
Since $\Delta$ increases 
with decreasing domain size, the average values over the multi-domain states 
also increase as the domain size becomes smaller.  

As the electric field is applied along [01], $P_{[10]}$ decreases to zero
for all three cases due to the switching of domains polarized
along [10] towards the [01] direction. As discussed earlier, 
$P_{[10]}$ reaches zero fastest for the smallest domain size, i.e.
$L_{0} \sim 0.06 \mu m$. The single domain $P_{[10]}$ remains zero, as
expected. In Fig. 4(b), we plot the average polarization along
the [01] direction, $P_{[01]}$,  as a function of the applied field $E_{[01]}$.
Since [01] is a polar
direction, $P_{[01]}$ grows for all the cases. Here also, the case
with the smallest domain size reaches the saturation value
the fastest. $P_{[01]}$ for the single domain varies only slightly
with the applied field as there is no domain switching for that
case.

To study the electromechanical behavior, we have also computed the
variation of the strains with the applied electric field. To
evaluate the contribution of the applied electric field to the
strain, we subtract off the zero field strain. Figures 5(a), 5(b)
and 5(c) show the behavior of average strain components  $\langle 
\eta_{xx}(E_{[01]})\rangle -\langle \eta_{xx}(E_{[01]}=0) \rangle$, $\langle
\eta_{yy}(E_{[01]}) \rangle-\langle \eta_{yy}(E_{[01]}=0)\rangle$
and
$\langle\eta_{xy}(E_{[01]})\rangle-\langle\eta_{xy}(E_{[01]}=0)\rangle$
respectively. Here $\eta_{xx}=Q_{11}{P_x}^2+Q_{12}{P_y}^2$,
$\eta_{yy}=Q_{11}{P_y}^2+Q_{12}{P_x}^2$ and
$\eta_{xy}={Q_{44}}{P_x}{P_y}$. Since the multi-domain states
switch to a single domain state polarized along [01], shrinkage
along the transverse [10] direction is observed, as can be seen in
Fig. 5(a). The magnitude of this transverse strain is almost the
same for all the multi-domain states, although the field required
to establish the single domain state increases as the domain size
is increased. The corresponding single domain undergoes very small
shrinkage along the transverse direction as there is no domain
switching involved. Figure 5(b) shows the behavior of the
longitudinal strain along the direction of the applied field for
the three multi-domain states as well as the corresponding single
domain states. An extension along the [01] direction is observed
for all the cases. However, the multi-domain states generate much
larger strains in comparison to the single domain state. This is
due to the $90^{o}$ domain switching in the multi-domain states
that results in the extra strain. Figure 5(c) shows the average
shear the crystal undergoes during the evolution depicted in Figs. 1, 2 
and 3. It is clear that the magnitude of the shear depends on the
number of domain walls in the system. This is due to the fact that
the domain walls in the unpoled multi-domain states are sheared
relative to the bulk. Upon applying the electric field, these domain
walls disappear resulting in a net shear strain. Thus the average
shear depends on the number of domain walls. The positive shear strain 
for low fields (``overshoot") is due to the domain switching process. 
Since there are no domain walls in the single domain state, shear 
strain is zero for all values of the electric field, as can be 
observed in Fig. 5(c).

We have also studied the domain size dependence of the
longitudinal piezoelectric coefficient $d^{[01]}_{33}$. The piezoelectric
coefficients are calculated from the slope of $\langle
\eta_{yy}(E_{[01]})\rangle-\langle\eta_{yy}(E_{[01]}=0)\rangle$ vs. 
$E_{[01]}$ curve in Fig. 5(b). Figure 6 shows the behavior of
$d^{[01]}_{33}$ vs. $E_{[01]}$ for the three multi-domain cases along
with the analogous single domain case. The high values observed in the electric 
field range $0-10$ kV/cm are due to the switching of domains. 
To clearly show the behavior of piezoelectric constants in the 
low-field regime, we replot the data of this figure in the inset for 
$d^{[01]}_{33} < 1200$. The data in the inset shows that the low-field 
piezoelectric coefficients are enhanced as the domain size is decreased. For 
example, for the smallest domain size $L_{0} \sim 0.06 \mu m$, 
$d^{[01]}_{33} \sim 1100$ pC/N compared to 
$d^{[01]}_{33} \sim 210$ for $L_{0} \sim 0.3 \mu m$. 
In the large field regime ($E_{[01]} > 10$ kV/cm),  $d^{[01]}_{33}$ is nearly 
equal for all the cases as they all correspond to a poled single domain state.
\subsection{FIELD APPLIED ALONG A NON-POLAR DIRECTION}
In this section, we study the case when the configurations
depicted in Figs. 1(a), 2(a) and 3(a) are subjected to an electric
field along the [11] direction. This situation is a $2D$ analog of
the experiments by Wada and Tsurumi \cite{dmsize} where the electric
field was applied along the [111] direction to tetragonal multi-domain
single crystals of $BaTiO_3$.

Here, we apply a quasi-static electric field along the [11] direction. 
The field is applied in increments of $\Delta E_{[11]}=0.92$ kV/cm 
and the configurations are allowed to relax for $t^{*}=100$ time steps after each change. For this case, application of the field does not immediately result in the
creation of a single domain state along [11]. Instead, the
multi-domain structure remains stable and the polarization vectors
rotate until an electric field induced transition to a [11]
polarized rhombic state takes place. This situation is depicted in
Figs. 7, 8 and 9 corresponding to the multi-domain states with
domain sizes $L_{0} \sim 0.3 \mu m, 0.15 \mu m$ and $0.06 \mu m$, respectively. 
It is observed that the field induced transition occurs at a lower electric 
field as the domain size decreases. This result corroborates our earlier
conclusion that the domain walls help nucleate the field induced
transition \cite{rapl}. Thus, the larger the number of domain walls,
the smaller the field required to induce the transition. For example,
for the smallest domain
size $L_{0} \sim 0.06 \mu m$, the transition occurs at ${E_{[11]}}\sim
4.6$kV/cm whereas for the largest domain size $L_{0} \sim 0.3 \mu m$, the
transition occurs at ${E_{[11]}}\sim 24$ kV/cm, a more than $50 \%$ change. 

The evolution of the components of the average polarization for
the situations depicted in Figs. 7, 8 and 9 is plotted  in
Figs. 10(a) and 10(b). The response of a single domain state
with initial polarization $(0,0.26)$Cm$^{-2}$ is also shown. The
zero field components $P_{[10]}$ and $P_{[01]}$ of the average
polarizations for the multi-domain states in Figs. 7, 8 and 9
are non-zero due to the coexisting domains of $(0.26,\Delta)$Cm$^{-2}$
and $(\Delta,0.26)$Cm$^{-2}$. Since the polarization vectors rotate
towards the [11] direction, the evolution of $P_{[10]}$ and
$P_{[01]}$ is almost identical as both the polarization variants
exist in nearly equal proportion. The single domain on the other
hand starts from $(0,0.26)$Cm$^{-2}$ till it transforms to a rhombic
state $(0.21,0.21)$Cm$^{-2}$. Figure 10 also shows that the field
required to transform the multi-domain state to a rhombic phase
depends on the number of domain walls in the system. However, the
polarization components after the transition are same for all the
cases as eventually a single domain rhombic state is established.

Figures 11(a), 11(b) and 11(c) show the evolution of $\langle
\eta_{xx}(E_{[11]})\rangle-\langle\eta_{xx}(E_{[11]}=0)\rangle$, 
$\langle
\eta_{yy}(E_{[11]})\rangle-\langle\eta_{yy}(E_{[11]}=0)\rangle$
and $\langle
\eta_{xy}(E_{[11]})\rangle-\langle\eta_{xy}(E_{[11]}=0)\rangle$, 
respectively. The results of this figure can be understood in
terms of the electric field induced symmetry changes. Let us first
examine the results for the single domain state. The zero field
initial state $(0,0.26)$ Cm$^{-2}$ corresponds to a
rectangular symmetry whereas the final state
$(0.21,0.21)$Cm$^{-2}$ corresponds to a rhombic symmetry. This
symmetry change is achieved by a uniaxial shrinkage along [01] and
a uniaxial extension along [10], as can be inferred from Figs.  
11(a) and 11(b) (notice the sharp jump near the field induced
transition). The behavior of shear strain [shown in Fig. 11(c)]
is governed by the rotation undergone by the polarization vector.
In contrast, the zero field multi-domain states correspond to a
nearly square macroscopic symmetry due to the coexistence of two
polarization (rectangular) variants. Hence, the multi-domain evolutions 
of Figs. 7, 8 and 9 effectively correspond to electric field
induced square to rhombic transitions. The jump due to the field
induced transition occurs at smaller electric field values as the
domain size is decreased. The uniaxial strain the crystal undergoes after 
the transition is almost the same along the [10] and [01] directions.
Interestingly, the saturation value of the strains is essentially the
same for all the three multi-domain evolutions. The magnitude of
shear strains after the transition, on the other hand, depends on
the domain size (or the number of domain walls) in the initial
state. As seen in Fig. 11(c), the amount of shear experienced by the crystal is
the largest for the case with $L_{0} \sim 0.3 \mu m$ and the smallest for the
$L_{0} \sim 0.06 \mu m$ case. Pre-existing shear strains at the domain
walls limit the total shear experienced by the multi-domain crystals
and thus the larger the number of domain walls in the initial
state, the smaller the shear strains produced.

Figure 12 depicts the behavior of the longitudinal piezoelectric
coefficients $d^{[11]}_{33}$ for the three multi-domain states as
well as the analogous single domain situation. The quantity
$d^{[11]}_{33}$ is calculated from the slope of the longitudinal
strain resolved along the [11] direction vs. $E_{[11]}$ curve. The
resolved strain is calculated as
$\langle\eta_{[11]}(E_{[11]})\rangle$
-$\langle\eta_{[11]}(E_{[11]}=0)\rangle$, where $\eta_{[11]}$ is
given by 
\begin{equation}
\eta_{[11]}={{1}\over{2}}({\eta_{xx}+\eta_{yy}+\eta_{xy}}).
\end{equation}
It is clear that the low-field piezoelectric coefficients for 
the smallest domain size $L_{0}\sim 0.06 \mu m$ are more enhanced 
compared to the domain patterns with $L_{0} \sim 0.15 \mu m$ and $0.3 \mu m$. 
The low field piezoelectric coefficients for the coarser domain patterns 
are not much higher than the single domain coefficients, consistent with 
recent experiments \cite{dragan}.
We believe that the 
enhancement is related to the response of unit cells in the 
domain wall regions  as such regions become bigger as the domain size becomes 
smaller. 
\section{SUMMARY AND DISCUSSIONS}
We have used a Ginzburg-Landau formalism to study
the domain size dependence of the piezoelectric properties. The
present work is inspired by the recent experiments of Wada and
Tsurumi \cite{dmsize} on domain engineered $BaTiO_{3}$ single crystals 
where the effect of the size of non-$180^{o}$ domains on the piezoelectric
constants was studied. In our model calculation, we solved the $2D$
time-dependent-Ginzburg-Landau equations \cite{rapl} with biased
initial conditions (the free energy parameters for
$BaTiO_{3}$ were chosen from Ref. \onlinecite{bell}) to create
three different multi-domain states with different domain widths.

Two different directions of the applied field
were considered.
In the first case, the multi-domain states were subjected to an
electric field along the [01] direction, which is one of the four
polar directions. The multi-domain states switched to single
domain states polarized along the [01] direction, with the state  
having the largest number of domain walls switching at the lowest
electric field. The
multi-domain state with the smallest domain size also exhibited
the largest value of the longitudinal piezoelectric constant
$d^{[01]}_{33}$. This enhancement of the piezoelectric coefficient as
the domain size is decreased reflects the metastability of the
multi-domain states which can become easily switchable as the
number of domain walls is increased. However,
this enhancement of the piezoelectric constant may not be very
useful in practical applications as the multi-domain states are
not stable over a large range of electric fields. 

We also considered the case where an electric field along the [11]
direction is applied to the same multi-domain states. This
situation is analogous to the experiments of Wada and Tsurumi
\cite{dmsize} who studied the piezoelectric properties of
tetragonal domain engineered $BaTiO_{3}$ single crystals under an
electric field along [111]. For this case, we found that the
multi-domain states remain stable until a field induced
rectangular to rhombic transition takes place. Interestingly, we
found that the transition occurs at smaller fields as the domain
size is decreased. Since proximity to the field induced transition
enhances the piezoelectric constants, the low-field piezoelectric
constant for the smallest size simulated by us is found to be
significantly higher than that for the single crystal and multi-domains
with bigger domain sizes. Thus, the role of the domain walls in
nucleating a field induced transition may be the cause of the
enhanced piezoelectricity in small sized engineered domains
observed by Wada and Tsurumi \cite{dmsize}. This enhancement may
be used in practical applications, provided the field is not too
close to the field induced transition.
\section{Acknowledgment} 
This work was supported by the U.S. Department of Energy and NSF 
Grant No. DGE-9987589.

\newpage
Figure Captions:
\vskip 0.4truecm 
Figure 1: Evolution of  domains for an electric field applied along the 
[01] direction to the multi-domain state with domain size $L_{0} \sim 0.3 
\mu m$. The corresponding electric field levels are indicated at 
the top of each snapshot.

Figure 2: Evolution of  domains for an electric field applied along the 
[01] direction to the multi-domain state with domain size $L_{0} \sim 0.15 
\mu m$. The corresponding electric field levels are indicated at 
the top of each snapshot.

Figure 3: Evolution of  domains for an electric field applied along the 
[01] direction to the multi-domain state with domain size $L_{0} \sim 0.06 
\mu m$. The corresponding electric field levels are indicated at 
the top of each snapshot.

Figure 4: Evolution of average polarizations $P_{[10]}$ (Fig. 4a) and $P_{[01]}$ 
(Fig. 4b) with the applied field $E_{[01]}$. The lines with circles correspond 
to the multi-domain state of Fig. 1, lines with crosses correspond to the 
multi-domain state of Fig. 2 and lines with squares correspond to multi-domain 
state of Fig. 3. Solid lines correspond to the  single domain state.

Figure 5: Evolution of average strains $\langle\eta_{xx}(E_{[01]})\rangle 
-\langle\eta_{xx}(E_{[01]}=0)\rangle$ (Fig. 5a), 
$\langle\eta_{yy}(E_{[01]})\rangle-\langle\eta_{yy}(E_{[01]}=0)\rangle$ 
(Fig. 5b) and 
$\langle\eta_{xy}(E_{[01]})\rangle-\langle\eta_{xy}(E_{[01]}=0)\rangle$ (Fig. 5c) 
with the applied field $E_{[01]}$. The lines with circles correspond to the 
multi-domain state of Fig. 1, lines with crosses correspond to the multi-domain 
state of Fig. 2 and lines with squares correspond to multi-domain state of Fig. 
3. Solid lines correspond to the  single domain state.

Figure 6: Variation of $d^{[01]}_{33}$ (the longitudinal piezoelectric constant 
along  [01]) with $E_{[01]}$.  The lines with circles correspond to the 
multi-domain state of Fig. 1, lines with crosses correspond to the multi-domain 
state of Fig. 2 and lines with squares correspond to multi-domain state of Fig. 
3. Solid lines correspond to the  single domain state. The inset plots the data 
in the range $d^{[01]}_{33} < 1200$ pC/N to show the low-field behavior.

Figure 7: Evolution of  domains for an electric field applied along the [11] 
direction to the multi-domain state with domain size $L_{0} \sim 0.3 \mu m$. 
The corresponding electric field levels are indicated at the top of each snapshot.

Figure 8: Evolution of  domains for an electric field applied along the [11] 
direction to the multi-domain state with domain size $L_{0} \sim 0.15 \mu m$. 
The corresponding electric field levels are indicated at the top of each snapshot.

Figure 9: Evolution of  domains for an electric field applied along the [11] 
direction to the multi-domain state with domain size $L_{0} \sim 0.06 \mu m$. 
The corresponding electric field levels are indicated at the top of each snapshot.

Figure 10: Evolution of average polarizations $P_{[10]}$ (Fig. 10a) and 
$P_{[01]}$ (Fig. 10b) with the applied field $E_{[11]}$. The lines with 
circles correspond to the multi-domain state of Fig. 7, lines with crosses 
correspond to the multi-domain state of Fig. 8 and lines with squares 
correspond to the multi-domain state of Fig. 9. Solid lines correspond to the  
single domain state.

Figure 11: Evolution of average strains 
$\langle\eta_{xx}(E_{[11]})\rangle-\langle\eta_{xx}(E_{[11]}=0)\rangle$ (Fig. 
11a), 
$\langle\eta_{yy}(E_{[11]})\rangle-\langle\eta_{yy}(E_{[11]}=0)\rangle$ (Fig. 11b) and 
$\langle\eta_{xy}(E_{[11]})\rangle-\langle\eta_{xy}(E_{[11]}=0)\rangle$ (Fig. 11c) with the applied field $E_{[11]}$. 
The lines with 
circles correspond to the multi-domain state of Fig. 7, lines with crosses 
correspond to the multi-domain state of Fig. 8 and lines with squares 
correspond to the multi-domain state of Fig. 9. Solid lines correspond to the  
single domain state.

Figure 12: Variation of $d^{[11]}_{33}$ (the longitudinal piezoelectric 
constant along  [11]) with $E_{[11]}$.  The lines with circles correspond to 
the multi-domain state of Fig. 7, lines with crosses correspond to the 
multi-domain state of Fig. 8 and lines with squares correspond to the 
multi-domain state of Fig. 9. Solid lines correspond to the single domain state.
\end{document}